\documentclass[runningheads]{svmult}
\usepackage{makeidx}   
\usepackage{graphicx}  
\usepackage{subeqnar}  
\usepackage{multicol}  
\usepackage{cropmark} 
\usepackage{physprbb}  
\begin{document}
\title*{Hot and Cold Dark Matter Search
\protect\newline with GENIUS
\thanks{Talk presented by Laura Baudis}}
\toctitle{Hot and Cold Dark Matter Search
\protect\newline with GENIUS}
%
%
\titlerunning{Hot and Cold Dark Matter Search with GENIUS}
%
\author{Laura Baudis
\and Alexander Dietz
\and Gerd Heusser
\and Hans Volker Klapdor-Kleingrothaus\thanks{Spokesman of the GENIUS collaboration}
\and Bela Majorovits
\and Herbert Strecker}

\authorrunning{Laura Baudis et al.}
%
%
\institute{Max--Planck--Institut f\"ur Kernphysik, Heidelberg,
  Germany}

\maketitle              

\begin{abstract}
GENIUS is a proposal for a large volume detector to search 
for rare events. An array of 40--400 'naked' HPGe detectors will 
be operated in a tank filled with ultra-pure liquid nitrogen.
After a description of performed technical studies of detector
operation in liquid nitrogen and of Monte Carlo simulations of  
expected background components, the potential of GENIUS 
for detecting WIMP dark matter, the neutrinoless double beta decay in
$^{76}$Ge and low-energy solar neutrinos is discussed.
\end{abstract}

\section{Introduction}

GENIUS (GErmanium in liquid NItrogen Underground Setup) is a proposal
for operating a large amount of 'naked' Ge detectors in liquid
nitrogen to search for rare events such as WIMP-nucleus scattering,
neutrinoless double beta decay and solar neutrino interactions, 
with a much increased sensitivity relative to existing experiments
 \cite{ringb,nim_genius,prop_genius}.
By removing (almost) all materials from the immediate vicinity  of 
the Ge-crystals, their absolute background can be considerably
decreased with respect to conventionally operated detectors. 
The liquid nitrogen acts both as a cooling medium and as a shield
against external radioactivity. The proposed scale of the experiment 
is a nitrogen tank of about 12\,m diameter and 12\,m height 
with 100\,kg of natural Ge and 1 ton of enriched $^{76}$Ge in 
the dark matter and double beta decay versions, respectively,
suspended in its center. 

To cover large parts of the MSSM parameter space, relevant for the
detection of neutralinos as  dark matter candidates \cite{jkg96,bedny1}, 
a maximum background
level of 10$^{-2}$ counts/(kg\,y\,keV) in the energy region below 
100\,keV has  to be achieved. In the double beta decay 
region (Q-value = 2038.56\,keV) a background of 0.3~events/(t\,y\,keV)
is needed in order to test the effective Majorana neutrino mass down
to 0.01\,eV.
This implies a very large background reduction in comparison to our recent best results 
\cite{prd,prl}  with the Heidelberg--Moscow experiment.

\section{Experimental studies and background considerations}

To demonstrate the feasibility of operating Ge detectors in liquid nitrogen
we performed three experiments in the Heidelberg low level
laboratory \cite{nim_genius,hellmig97,diss}.
The goal was to look for possible interferences between 
two or more naked Ge crystals, to test different cable lengths between
FETs and crystals and to design and test preliminary holder systems.
For crystal masses between 300--400\,g   we achieved  
energy resolutions of about 1.0\,keV at 300\,keV and thresholds of
about 2\,keV. 
No microphonic events due to nitrogen boiling beyond 2\,keV could be detected. 
Also, we couldn't observe any cross talk using only p--type
detectors (same polarity for the HV-bias), 
since cross talk signals have the wrong  polarity and are filtered by the amplifier.
Concluding, the performance of the Ge detectors is as good (or even
better) as for conventionally operated crystals, even with 6\,m cable
lengths between crystal and FET. 

For an estimation of the expected overall background in both low and high 
energy regions, we performed detailed Monte Carlo
simulations of all the relevant background sources.
The sources of background can be divided into external and internal ones.
External background is generated by events originating from outside
the liquid shield, such as photons and neutrons from the Gran Sasso
rock, muon interactions and muon induced activities.
Internal background arises from residual impurities in the liquid
nitrogen, in the steel vessel, in the crystal holder system, in the Ge crystals
themselves and from activation of both liquid nitrogen and Ge crystals
at the Earths surface.
For the simulation of muon showers, the external photon flux
and the radioactive decay chains we used the GEANT3.21 package 
\cite{geant} extended for nuclear decays \cite{mueller}.

The simulated geometry consisted of a cylindrical nitrogen vessel 
of 12\,m in diameter and 12\,m in height,  
surrounded by a 2\,m thick polyethylene-foam isolation, which 
is held by two 2\,mm thick steel layers. The crystals were
held by a holder system of high molecular polyethylene and 
positioned in the tank centre.

\subsubsection*{External background}

We simulated the measured photon \cite{arpesella92}, neutron \cite{arp} 
and muon \cite{macro} fluxes in the Gran Sasso underground laboratory.
The underlying assumptions were a 12\,m$\times$12\,m nitrogen shield, a 2\,m 
thick boron loaded polyethylene foam isolation and a muon veto with a 96\% 
efficiency on top of the tank \cite{nim_genius}.
The resulting count rates  for both low and high energy regions are
shown in Table \ref{sim_1}. 

\begin{table}
\caption{Resulting count rates for the simulation of the gamma, 
neutron and muon fluxes measured in the Gran Sasso laboratory, in the 
energy regions  between 11\,keV--100\,keV and 
2000\,keV - 2080\,keV.} 
\begin{center}
\renewcommand{\arraystretch}{1.4}
\setlength\tabcolsep{7pt}
\begin{tabular}{lll}
\hline\noalign{\smallskip}
Component & Count rate (11-100\,keV) & Count rate (2000-2080\,keV)\\
                    &  [events/(kg\,y\,keV)] &[events/(t\,y\,keV)] \\
\hline
\noalign{\smallskip}
 gammas      &  4$\times$10$^{-3}$ & 2$\times$10$^{-1}$\\
                neutrons    &  4$\times$10$^{-4}$ & 6$\times$10$^{-3}$\\
                muon showers &  2$\times$10$^{-3}$ & 2$\times$10$^{-2}$\\
                $\mu$ $\rightarrow$ n, $^{71}$Ge, $^{77}$Ge   &  1$\times$10$^{-3}$ 
&1.2$\times$10$^{-2}$\\
                $\mu$ $\rightarrow$ caption & $<<$1$\times$10$^{-4}$ &$<<$1$\times$10$^{-4}$\\ 
\hline 
\end{tabular}
\end{center}
\label{sim_1}
\end{table}

The anticoincidence of the 40 (400) Ge-detectors further reduces 
the effect of muon showers by a factor of 5 (100).
Besides muon showers, we considered muon induced nuclear 
disintegration and interactions due to secondary neutrons generated in the above
reactions. 
Secondary neutron induced interactions in the liquid nitrogen, as well 
as negative muon capture and inelastic muon scattering reactions 
generate only a 
negligible contribution to the overall expected background rate (for 
details see \cite{nim_genius,diss}).
In germanium, two n-capture reactions are important (Table 
\ref{sim_1}):  $^{70}$Ge(n,$\gamma$)$^{71}$Ge  and 
$^{76}$Ge(n,$\gamma$)$^{77}$Ge.
$^{71}$Ge decays by EC (100\%) with T$_{1/2}$ = 11.43 d
and Q$_{\rm EC}$ = 229.4\,keV \cite{firestone}.
$^{77}$Ge decays by $\beta^-$-decay with T$_{1/2}$ = 11.3 h
and Q$_{\beta^-}$ = 2.7 MeV \cite{firestone}.
Because of their long half-lifes, these decays can not be discriminated by 
anticoincidence with the muon veto.

\subsubsection*{Internal background}

The assumed intrinsic impurity levels for the simulated materials
are listed in Table ~\ref{intr-assumpt}.

\begin{table}
\caption{ Assumed intrinsic impurity levels for the simulated 
detector components.}
\begin{center}
\renewcommand{\arraystretch}{1.4}
\setlength\tabcolsep{7pt}
\begin{tabular}{lll}
\hline\noalign{\smallskip}
Source & Radionuclide & Purity \\
\hline\noalign{\smallskip}
Nitrogen & $^{238}$U  & 3.5$\times$10$^{-16}$g/g \\
           & $^{232}$Th & 4.4$\times$10$^{-16}$g/g \\
           & $^{40}$K   & 1$\times$10$^{-15}$g/g \\
           & $^{222}$Rn &   3$\mu$Bq/m$^3$\\ 
\hline\noalign{\smallskip}
Ge crystals &  $^{238}$U  & $<$1.8$\times$10$^{-15}$g/g \\
           & $^{232}$Th & $<$5.7$\times$10$^{-15}$g/g \\
\hline  \noalign{\smallskip}   
Holder system  & U/Th       & 1$\times$10$^{-12}$g/g \\
\hline\noalign{\smallskip}
Steel vessel  & U/Th       & 5$\times$10$^{-9}$g/g \\
\hline
\end{tabular}
\end{center}
\label{intr-assumpt}
\end{table}

The values assumed for the $^{238}$U and $^{232}$Th decay chains in 
liquid nitrogen have already been
reached by BOREXINO \cite{borexino} for their liquid scintillator. 
Due to the very high cleaning efficiency of fractional distillation, 
it is conservative to assume
that these requirements will also be fulfilled for liquid nitrogen.
The $^{222}$Rn contamination of freshly
produced liquid nitrogen was recently measured to be 325\,$\mu$Bq/m$^3$ 
\cite{rau}. Here, an additional underground storage time of 1 month was assumed. 
This level could be maintained if the evaporated nitrogen is always
replaced by Rn-pure nitrogen, previously stored below ground.
Surface emanations are reduced to a negligible level for cooled
surfaces in direct contact with the liquid nitrogen.
The intrinsic impurity levels in Ge crystals are 
conservative upper limits from measurements with the detectors of the
Heidelberg--Moscow experiment. 
We see a clear $\alpha$--peak in two of the enriched detectors at 5.305
MeV, and an indication for the same peak in two other detectors. It
originates from the decay of $^{210}$Po (which decays with 99\%
through an $\alpha$--decay to $^{206}$Pb) and is a sign for a $^{210}$Pb
contamination of the detectors. 
The assumed values for polyethylene were 
reached by the SNO experiment
\cite{sno}, for an acrylic material.
The impurity level in  steel  
has been measured by BOREXINO \cite{borexprop}. 

The results of the simulation of the intrinsic background components is given in Table
\ref{sim-2}. Not included is the contribution from the Ge-crystals 
 (about 1$\times$10$^{-2}$ events/(kg\,y\,keV) below 100\,keV)
since it is very unlikely that the
contaminations are intrinsic of the crystals. Most probably they are
on the crystal surfaces at the inner contacts. Besides that, it is
very unlikely that $^{210}$Pb is in equilibrium with $^{238}$U, as we
assumed.
     However, for GENIUS, special attention will have to be paid in order
to avoid such surface contaminations of the 'naked' crystals.

\begin{table}
\caption{Low and high energy count rates from the intrinsic impurities 
of the simulated detector components.}
\begin{center}
\renewcommand{\arraystretch}{1.4}
\setlength\tabcolsep{7pt}
\begin{tabular}{llll}
\hline
Source & Component & Count rate (11-100\,keV) & Count rate (2000-2080\,keV)\\
&                    &  [events/(kg\,y\,keV)] &[events/(t\,y\,keV)] \\
\hline
Nitrogen & $^{238}$U   &  7$\times$10$^{-4}$ & 2$\times$10$^{-4}$\\
intrinsic  & $^{232}$Th  &  4$\times$10$^{-4}$ & 7$\times$10$^{-4}$\\
           & $^{40}$K    &  1$\times$10$^{-4}$  &  ---\\
           & $^{222}$Rn  &  3$\times$10$^{-4}$ &   $<$10$^{-4}$\\
\hline
Steel vessel   & U/Th        &  1.5$\times$10$^{-5}$ &3$\times$10$^{-3}$\\
\hline
Holder  & U/Th        &  8$\times$10$^{-4}$ &1$\times$10$^{-4}$\\
\hline 
\end{tabular}
\end{center}
\label{sim-2}
\end{table}

We have estimated the cosmogenic production rates of radioisotopes 
in the Ge--crystals with the $\Sigma$ programme \cite{JensB}.
Assuming a production and transportation time of 10 days at sea level
for the Ge--detectors, and a deactivation time of three\,years,
we obtain the radioisotope concentrations listed in Table
\ref{ge_cosmo} (for $^{68}$Ge the saturation activity was assumed).
All other produced radionuclides have much smaller activities due to their shorter
half lifes. 

The count rate below 11\,keV is dominated by X--rays from 
the decays of $^{68}$Ge, $^{49}$V and     
$^{55}$Fe (see Table \ref{ge_cosmo}).        
Due to their strong contribution, the energy threshold of GENIUS would 
 be 11\,keV, which is still acceptable (as can be seen from 
Fig. \ref{limits}).

\begin{table}
\caption{Cosmogenically produced isotopes in the Ge crystals for an
  exposure at sea level of 10 days and a subsequent deep underground
  storage of 3\,years (for $^{68}$Ge the saturation activity was assumed).}
\begin{center}
\renewcommand{\arraystretch}{1.4}
\setlength\tabcolsep{5pt}
\begin{tabular}{llll}
\hline\noalign{\smallskip}
Isotope & Decay, T$_{1/2}$ & Energy deposition in the cystal [keV]& 
A [$\mu$Bq kg$^{-1}$]\\
\hline\noalign{\smallskip}
$^{49}$V   &  EC, 330 d    & E$_K$(Ti)= 5, no $\gamma$ & 0.17\\
$^{54}$Mn  &  EC, 312.2 d  & E$_{\gamma}$= 840.8    & 0.20\\
$^{55}$Fe  &  EC, 2.73 a   & E$_K$(Mn)= 6.5, no $\gamma$    & 0.31\\
$^{57}$Co  &  EC, 271.3 d  & E$_{\gamma}$= 136.5  & 0.18\\
$^{60}$Co  &  $\beta ^-$, 5.27 a & E$_{\beta^-}$= 318, E$_{\gamma}$=
1173.2,1332.5  &0.18\\
$^{63}$Ni & $\beta^-$, 100.1 a & E$_{\beta^-}$= 66.95, no $\gamma$ & 0.01\\
$^{65}$Zn & EC, 244.3 d & E$_{\gamma}$=1125.2  & 1.14\\
$^{68}$Ge & EC, 288 d & E$_K$(Ga)=10.37, $^{68}$Ga decay &101\\
\hline
\end{tabular}
\end{center}
\label{ge_cosmo}
\end{table}

\begin{figure}
\begin{center}
\includegraphics[width=0.8\textwidth]{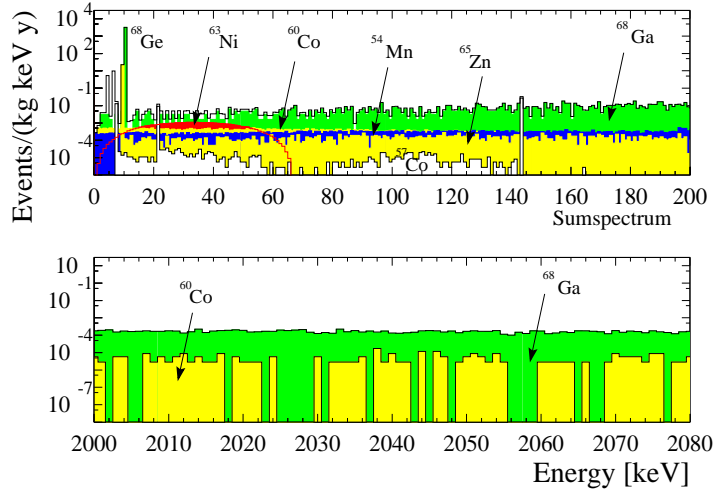}
\end{center}
\caption{Background originating from cosmic activation of the Ge
  crystals at sea level with 10 days exposure and 3\,years deactivation.} 
\label{cosmo}
\end{figure}

Between 11\,keV and 70\,keV the contribution from $^{63}$Ni dominates
due to the low Q--value (66.95\,keV) of the $\beta^-$--decay. 
Figure \ref{cosmo} shows the sum and the single contributions from the
different isotopes.
$^{68}$Ge plays a special role. Since it can not be
extracted by zone melting like all other, non--germanium isotopes,
the starting activity would be in equilibrium with the production
rate. With a half--life of 288 d it would by far dominate the other background
components. A solution could be to process the germanium ore directly
below ground or to use high purity germanium which
has already been stored for several\,years in an underground laboratory.    

The sum of all contributions from the cosmogenic activation of the Ge
crystals amounts to 5.2$\times$10$^{-2}$ counts/(kg\,y\,keV) between
11--100\,keV, and to 3.6$\times$10$^{-2}$ counts/(t\,y\,keV) between
2000--2080\,keV (a de-enrichment factor of 500 for the $^{70}$Ge in the enriched
$^{76}$Ge material was assumed). 
Since this will be the dominant background component in the
low energy region, special attention to short crystal exposure times at
sea level is essential.

The relevant cosmogenic isotopes produced in the nitrogen
during its transportation at sea level are given in Table 
\ref{cosmo-n} along with their decay modes and energies.
Their estimated activities for a realistic 10 days exposure to 
a sea level neutron flux  of
8.2$\times$10$^{-3}$cm$^{-2}$s$^{-1}$ 
(for neutron energies between 80 MeV and 300 MeV ) \cite{allkofer} 
and  to a muon flux of 1.6$\times$10$^7$m$^{-2}$d$^{-1}$ 
along with the induced background rates in the low energy region are also shown.

\begin{table}
\caption{Cosmogenically produced radionuclides in liquid nitrogen at sea
  level, estimated activities for a 10\,d exposure and  induced background rates.}
\begin{center}
\renewcommand{\arraystretch}{1.4}
\setlength\tabcolsep{5pt}
\begin{tabular}{lllll}
\hline\noalign{\smallskip}
Isotope & T$_{1/2}$, Decay & Energy & Activity & Rate (11-100\,keV)\\
        &                 &        &[Bq/g]    &  [ev./(kg\,y\,keV)]\\ 
\hline\noalign{\smallskip}
$^{3}$H & 12.35 y, $\beta^-$ (100\%)& E$_{{\beta}^-}$ = 18.6 keV& 3.8$\times$10$^{-8}$ & ---\\
$^{7}$Be & 53.29 d, EC, $\gamma$ (10\%) & E$_{\gamma}$ = 477.61 keV&3.7$\times$10$^{-9}$ & 8$\times$10$^{-3}$ \\
$^{10}$Be & 1.6$\times$10$^6$ y,$\beta^-$ (100\%)& E$_{{\beta}^-}$ = 555 keV & 8.4$\times$10$^{-15}$ & negligible\\
$^{14}$C & 5.7$\times$10$^4$ y,$\beta^-$ (100\%)& E$_{{\beta}^-}$ = 156 keV & 1.4$\times$10$^{-4}$ & 1$\times$10$^{-4}$ \\
\hline
\end{tabular}
\end{center}
\label{cosmo-n}
\end{table}

No events of tritium decay were detected in the Ge-crystals, 
mainly due to the absorption in the dead layer of the p-type 
Ge detectors. For $^{7}$Be, the additional assumption of one year of 
deactivation in Gran Sasso was made. Moreover, it can be
expected, that a large fraction of $^7$Be is removed
from the liquid nitrogen at the cleaning process for Rn.
For the production of $^{10}$Be and $^{14}$C both neutron and muon
capture induced channels were considered.

Summing up the background contributions discussed so far,
the mean count rate  in the low energy 
region amounts to about 6$\times$10$^{-2}$ events/(kg\,y\,keV) and 
to 2.8$\times$10$^{-1}$ events/(t\,y\,keV) in the region relevant for the 
0$\nu\beta\beta$-decay.

\begin{figure}
\begin{center}
\includegraphics[width=.7\textwidth]{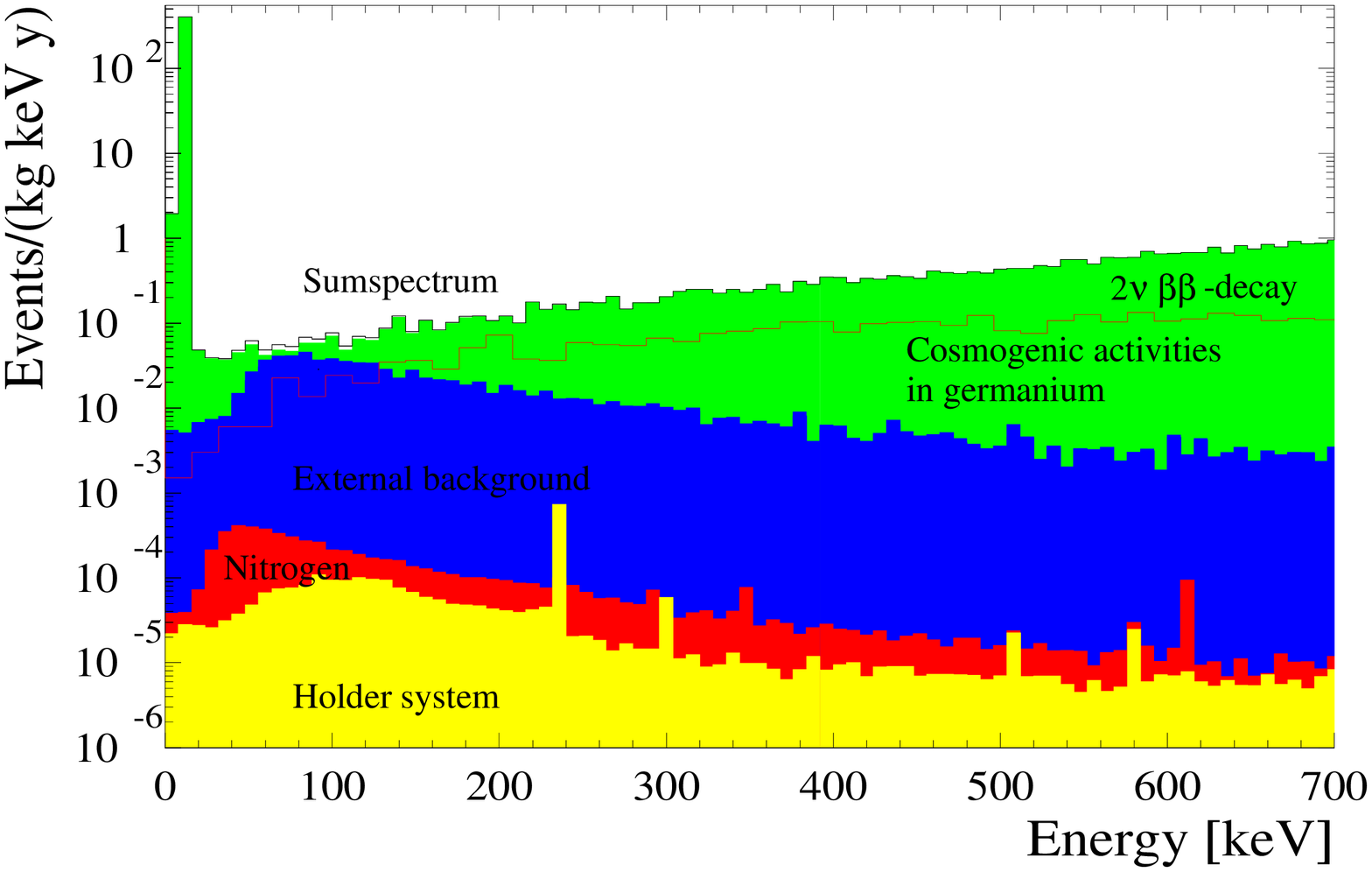}
\includegraphics[width=.7\textwidth]{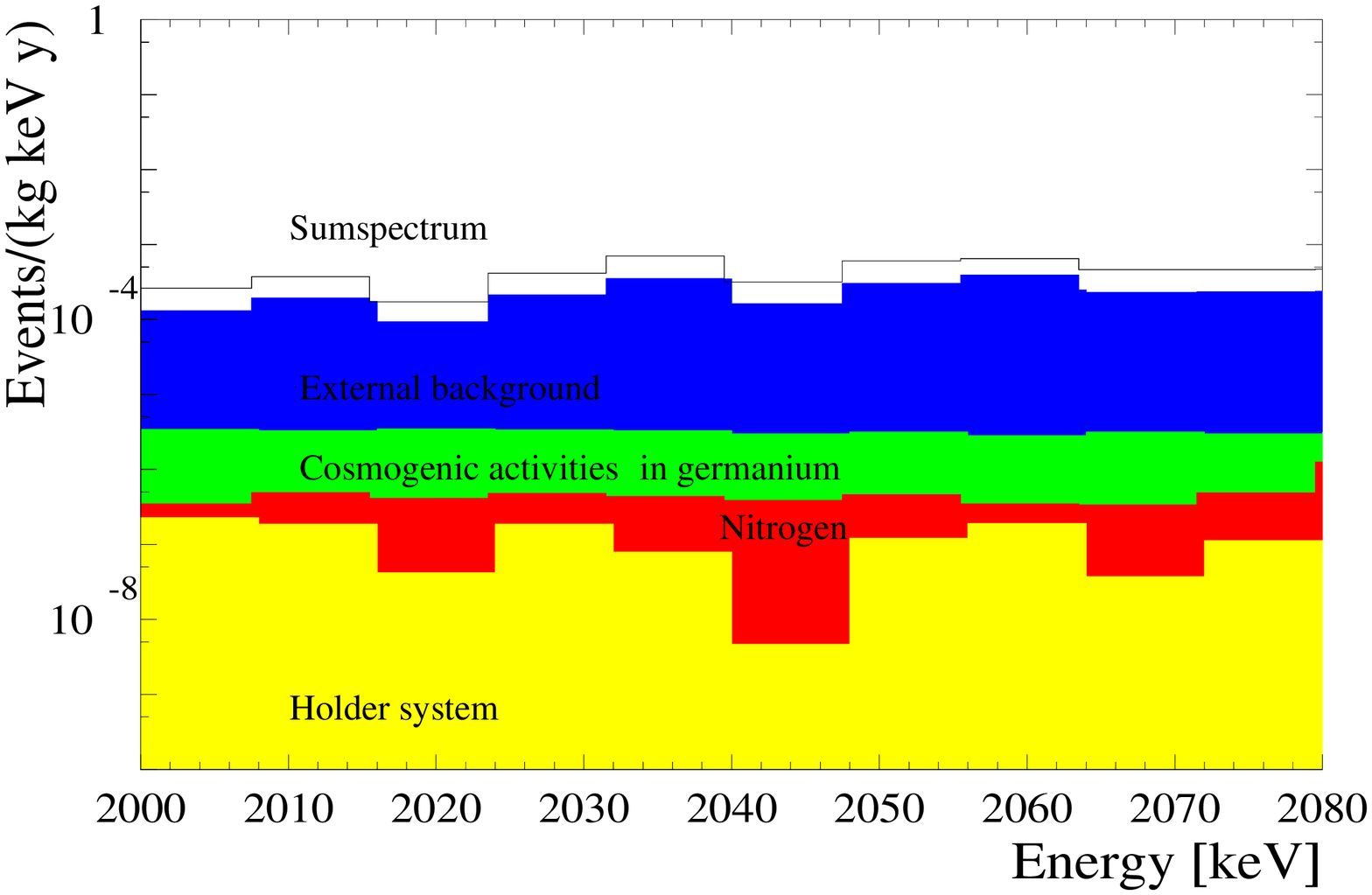}
\end{center}
\caption{Simulated spectra of the dominant background sources for
a nitrogen tank of 12\,m diameter. Three years of storage below ground 
for the Ge detectors were assumed.
Shown are the contributions from the detector holder
system, from the intrinsic nitrogen contamination,
from the external natural radioactivity and from the cosmogenic
activation of the Ge detectors.
The solid line represents the sum spectrum of all the simulated
components.}
\label{specall}
\end{figure}

In Fig. \ref{specall} the spectra of individual contributions 
and the summed up total background spectrum are shown (after three
years of storage of the Ge detectors below ground). 
The low energy spectrum is dominated by events
originating from the cosmogenic activation of the Ge crystals at the
Earths surface. Another two years of storage below ground, or
production of the detectors in an underground facility would
significantly reduce this contribution.  
For the high-energy region, the results of the simulations are 
comparable to the aim of  0.3\,counts/(t\,y\,keV). The 
background spectrum is dominated by the contribution of external 
gammas (which again reveals the importance of a 12\,m$\oslash$ LiN 
shield) followed by the contribution of the cosmogenic $^{60}$Co.

\section{Potential of GENIUS to search for rare events}

\subsubsection*{WIMP dark matter}

\begin{figure}
\begin{center} 
\includegraphics[width=.8\textwidth]{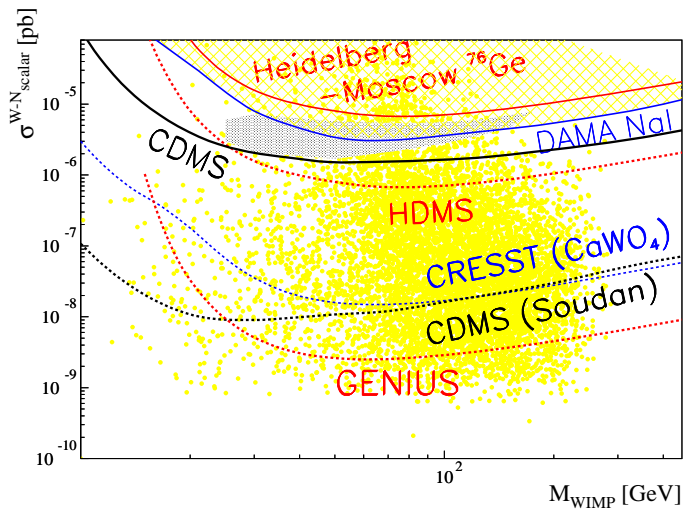}
\caption{
WIMP-nucleon cross section limits as a function of the WIMP
  mass for spin-independent interactions. 
The hatched region is excluded by the Heidelberg-Moscow
 \cite{prd} and the DAMA experiment \cite{dama}, the plain black
 curve is the new limit of the CDMS experiment \cite{rick2000}.
 The dashed lines are 
  expectations for recently started or future experiments, like HDMS
  \cite{hdms}, CRESST \cite{cresst}, CDMS (Soudan) \cite{cdms}
  and GENIUS \cite{prop_genius}. The
  filled contour represents  the 2$\sigma$ evidence region of the DAMA
  experiment \cite{dama3}.    
The experimental limits are compared to
expectations (scatter plot) for WIMP-neutralinos calculated in the
MSSM parameter space at the weak scale (without any GUT constraints)
under the assumption that all superpartner masses are lower than
300 GeV - 400 GeV \cite{vadim99}. }
\end{center} 
\label{limits}
\end{figure}

With 100\,kg of natural Ge and a background of 1$\times$10$^{-2}$ events/(kg\,y\,keV) 
in the energy region between 11--100\,keV, GENIUS could test a large 
part of the predicted MSSM parameter space for neutralinos as dark 
matter candidates.
Figure \ref{limits} shows a comparison of existing constraints and 
future sensitivities of cold dark matter experiments, together with the 
theoretical expectations for neutralino scattering rates 
\cite{vadim99}.
Even if the background would be higher than expected,
 GENIUS could easily cover the range of positive 
evidence for dark matter
 claimed by the DAMA experiment \cite{dama3}. 
It would be an independent test by using a different technology and 
only raw data, without any background subtraction.
Moreover, GENIUS would be the only experiment which 
could test DAMA directly, having a realistic 
chance to see the predicted  seasonal variation of the event rate in
100\,kg of detector material.

\subsubsection*{Neutrinoless double beta decay}

Neutrinoless double beta decay provides a powerful method for 
gaining informations about the absolute neutrino mass scale 
and a unique method of discerning between a Majorana and a Dirac neutrino. 
The current most stringent experimental limit on the effective  
Majorana neutrino mass, $\langle {\rm m} \rangle < $0.36 eV, comes 
from the Heidelberg-Moscow experiment \cite{prl,diss}.
For a significant step beyond this
limit, much higher source strengths and lower background levels are
needed.
This goal could be accomplished by the GENIUS experiment operating 
300--400
detectors made of enriched $^{76}$Ge (1 ton).
With  a background rate of 0.3\,events/(t\,y\,keV) in the energy region 
between 2000--2080\,keV, GENIUS would  reach 
the sensitivity of
$\langle {\rm m} \rangle < $ 0.01\,eV after one\,year of measuring
time.  This would have
striking influence on presently discussed neutrino mass scenarios
\cite{paes} and 
would allow a breakthrough into the multi 
TeV range for many beyond standard models \cite{ringb}. 
Already now the result from the Heidelberg-Moscow experiment
excludes simultaneous solutions for hot dark matter, the atmospheric
neutrino problem and the small mixing angle MSW solution of the solar
neutrino problem \cite{adh98}, as well as one of the two solutions for 
four-neutrino scenarios \cite{bil99}. GENIUS would determine the
mixing in partially or completely degenerate neutrino mass schemes 
with extreme accuracy, providing informations being complementary to
precision tests of cosmological parameters by MAP and Planck \cite{paes}. 
Besides that, it could test the LSND indication for neutrino
oscillations. For a detailed discussion of these topics we refer to
\cite{prop_genius,paes}.
Moreover, GENIUS would yield information on
supersymmetry (R-parity breaking, sneutrino mass), leptoquarks
(leptoquark -Higgs coupling or leptoquark mass), compositeness,
right-handed W boson mass, test of special relativity and equivalence
principle in the neutrino
sector, neutrino magnetic moment and others, competitive to
corresponding results from  future high-energy colliders \cite{prop_genius,kk1,kk2}.

\subsubsection*{Solar neutrinos}

The very low background aimed at by GENIUS in the low energy region, its 
energy threshold of about 11\,keV  and a 
target mass of at least  1 ton of natural (or enriched) Ge would open the possibility 
to look for pp- and $^{7}$Be solar neutrino interactions in real-time.
The detection reaction would be the elastic scattering process 
$\nu$ +  e$^- \rightarrow$ $\nu$ +  e$^-$.
The maximum electron recoil energy is 261\,keV for the pp-neutrinos and 665 
keV for the  $^7$Be-neutrinos \cite{bahc89}. 
The dominant part of the signal in GENIUS would be produced by 
pp-neutrinos (66\,\%) and the  $^7$Be-neutrinos (33\,\%) \cite{la_nu}.

A target mass of 1 ton (10 tons) of natural or enriched Ge corresponds
to about 3$\times$10$^{29}$ (3$\times$10$^{30}$) electrons.
Using the cross section for elastic neutrino-electron scattering 
from \cite{bahc89}
and the neutrino fluxes from \cite{Bah98c},
the expected number of events 
in the standard solar model (BP98 \cite{SSM}) can be estimated:\\

\noindent
R$_{pp}$ = 68.9 SNU = 1.8 events/day (18 events/day for 10 tons)\\
R$_{^7Be}$ = 28.5 SNU = 0.6 events/day (6 events/day for 10 tons),\\

\noindent
The event rates for full $\nu_e \rightarrow \nu_{\mu}$ conversion 
are 0.48 events/day for pp-neutrinos and 0.14 events/day for
$^7$Be-neutrinos for 1 ton of Ge and ten times higher for 10 tons. 
        
In order for GENIUS to be sensitive to the low-energy solar neutrino
flux, the background requirements would be more stringent than for the dark 
matter version.  
A nitrogen shield of 13\,m in diameter and 13\,m in height  is required.
Regarding the radiopurity of liquid nitrogen, the values reached at
present by the BOREXINO collaboration for their liquid scintillator
would be sufficient. More attention has to be paid to the cosmogenic
activation of the Ge crystals at the Earth surface. In case of one day 
exposure, five\,years of deactivation below ground are required.
The optimal solution would be to produce the detectors in an underground 
facility.

If the signal to background ratio (S/B) in GENIUS will be greater than 1, than the 
pp- and $^7$Be-neutrino flux can be measured by spectroscopic techniques
alone. 
If S/B $<$ 1,  one can make use of a solar signature in order to derive
the flux.
The eccentricity of the Earths orbit induces a seasonal variation
of about 7\% from maximum to minimum. Even if the number of background
events is not known, the background event rate and the signal event rate
can be extracted independently by fitting the event rate to the seasonal
variation. The only assumption is that the background is stable in time
and that enough statistics is available. 
In case of a day/night - variation of the solar neutrino flux,
GENIUS would be sensitive to the LOW MSW solution of the
solar neutrino problem.

\section{Summary and Outlook}

The capability of the GENIUS project to search for rare events 
such as WIMP-nucleus scattering, neutrinoless double beta decay and 
low energy solar neutrino interactions has been reviewed.
After presenting experimental studies which confirm 
the good performance of 'naked' Ge-crystals in liquid nitrogen, the 
background requirements were discussed in some detail.
The results achieved by Monte Carlo simulations for both low and high 
energy regions are promising.

Reaching the background level aimed at, the GENIUS project could bring
a large progress in the field of direct dark matter detection.
It could probe 
a relevant part of the SUSY--WIMP parameter space interesting for the detection of
neutralinos, thus possibly deciding whether or not neutralinos are the major
component of the dark matter in our Galaxy. 
In its double beta decay version, GENIUS could deliver important 
insights on the absolute neutrino mass scale, probing the 
effective Majorana neutrino mass down to 0.01 eV.
Last but not least,  it could be the first detector to detect the solar pp neutrinos
in real-time.

\end{document}